\documentclass[twocolumn,floatfix,showpacs,preprintnumbers,amsmath,amssymb,prb]{revtex4}
\usepackage{graphicx,tabularx}
\usepackage{amssymb}
\usepackage[mathcal]{euscript}
\newcommand{\abs}[1]{\left\vert#1\right\vert}

\newcommand{\limq}{\lim_{|\mathbf{q}|\to0}}
\newcommand{\bracet}[2]{\left<#1\vert#2\right>}

\newcommand{\bk}{\mathbf{k}}
\newcommand{\bq}{\mathbf{q}}

\begin{document}
\title{First-principles study of the optical properties of Mg$_x$Ti$_{(1-x)}$H$_2$}

\author{M.J. van Setten$^1$}
\author{S. Er$^2$}
\author{G. Brocks$^2$}
\author{R.A. de Groot$^{1,3}$}
\author{G.A. de Wijs$^1$}

\affiliation{$^1$Electronic Structure of Materials, Institute for Molecules and Materials, Faculty
of Science, Radboud University Nijmegen, Toernooiveld~1, 6525~ED Nijmegen, The~Netherlands}

\affiliation{$^2$Computational Materials Science, Faculty of Science and Technology and MESA+ Institute
for Nanotechnology, University of Twente, P.O.~Box~217, 7500~AE Enschede, The~Netherlands}

\affiliation{$^3$Laboratory of Solid State Chemistry, Zernike
Institute for Advanced Materials, Rijksuniversiteit Groningen,
Nijenborgh~4, 9747~AG Groningen, The Netherlands}

\date{\today}



\begin{abstract}
The optical and electronic properties of Mg-Ti hydrides are
studied using first-principles density functional theory.
Dielectric functions are calculated for Mg$_x$Ti$_{(1-x)}$H$_2$
with compositions $x=0.5$, 0.75, and 0.875. The structure is
that of fluorite TiH$_2$ where both Mg and Ti atoms reside at
the Ti positions of the lattice. In order to assess the effect
of randomness in the Mg and Ti occupations we consider both
highly ordered structures, modeled with simple unit cells of
minimal size, and models of random alloys. These are simulated
by super cells containing up to 64~formula units ($Z=64$). All
compositions and structural models turn out metallic, hence the
dielectric functions contain interband and intraband free
electron contributions. The former are calculated in the
independent particle random phase approximation. The latter are
modeled based upon the intraband plasma frequencies, which are
also calculated from first-principles. Only for the models of
the random alloys we obtain a black state, i.e.\ low reflection
and transmission in the energy range from 1 to 6~eV.
\end{abstract}

\maketitle
\section{Introduction}

Since the discovery of the switchable mirror YH$_x$ by Huiberts
{\it et al.}\cite{huiberts96} in 1996 several other metal
hydride systems that behave as switchable mirrors have been
discovered.\cite{huiberts96, vandersluis97, richardson01,
isidorsson02, lohstroh04, borsa06}. The metals are reflective,
but after hydrogenation become semiconductors and hence in most
cases become transparent. Especially when an alloy with high
hydrogen mobility is used and when applied as thin films the
optical switching can be fast, reversible and
robust.\cite{lohstroh07}

Recently, meta-stable thin films composed of various ratios of
magnesium and titanium have been shown to exhibit remarkable
optical properties which could be especially useful for smart
solar cell coatings and hydrogen sensor
applications.\cite{borsa06,slaman07} In the dehydrogenated
state the films are highly reflective. Upon hydrogenation they
become black, i.e.\ have a low reflection and high absorption
in the energy range of the solar spectrum. The structural and
electronic characteristics of this black state are a topic of
intensive research.\cite{borsa07,er07}

Obtaining structural data for these systems is difficult.
Single crystals cannot be grown as the compounds are
thermodynamically unstable. For a 7:1 ratio of magnesium to
titanium Kyoi {\it et~al.}, using a sample prepared at high
pressure, could determine the crystal structure.\cite{kyoi04}
It is similar to the fluorite TiH$_2$ structure. Notten and
co-workers (Refs.~\onlinecite{niessen05, vermeulen06}) and
Borsa {\it et al.}\cite{borsa06,borsa07} have shown that for
higher titanium content ($x < 0.8$) the structure is
fluorite-like as well. At lower lower titanium concentrations a
rutile ($\alpha$-MgH$_2$ like) structure is found.
Interestingly, the kinetics of the hydrogen ab/desorption
reactions are much faster in the fluorite structures than in
the rutile structures.\cite{vermeulen06-2} The equality of the
molar volume of TiH$_2$ and Mg has been used to explain the
structural stability of the meta-stable
``phases''.\cite{borsa07} Calculations using density functional
theory (DFT) find the same composition dependence of the
relative stability of the rutile and fluorite
structures.\cite{er07}

The origin of the ``black state'' is not understood. Its
explanation is, of course, intimately related to the electronic
structure of the hydride. In experiments, usually less than two
hydrogen atoms per metal atom can be reversibly absorbed and
desorbed.\cite{borsa07} Electrochemically 1.7 hydrogen atoms
per metal atom can be stored reversibly.\cite{vermeulen06-2}
Moreover, the crystal structure of Mg$_7$TiH$_x$ as determined
by Kyoi {\it et al.}\cite{kyoi04} is estimated to contain 1.6
hydrogen atoms per metal atom. All this suggests that maximally
two metal electrons per metal atom can be transferred to
hydrogen.\cite{borsa07} Hence the Ti atoms remain in an
open-shell configuration with at least two $d$-electrons left.
The above reasoning has been confirmed by DFT calculations on
crystalline Mg$_x$Ti$_{(1-x)}$H$_2$ structures.\cite{er07} The
calculated densities of states show predominant hydrogen-$s$
character below the Fermi level, which is typical for metal
hydrides, and titanium-$d$ states at the Fermi level. These are
likely to form metallic bands, so one expects
Mg$_x$Ti$_{1-x}$H$_2$ to be reflective instead of black.
Moreover, experimental investigation of the electrical
transport properties reveal high resistivity
1.32--1.9~m$\Omega$cm, with a logarithmic, hence non-metallic,
temperature dependence.\cite{borsa07} In order to explain these
results the formation of a so-called coherent crystal structure
was proposed,\cite{borsa07} wherein regions of insulating
MgH$_2$ and metallic TiH$_2$ coexist.

In an effort to advance the understanding of the black state,
in this paper we report a computational study of the optical
and electronic properties of Mg$_x$Ti$_{(1-x)}$H$_2$ for $x =
0.5$, 0.75 and 0.875. We employ both simple (i.e.\ minimal)
unit cells and large super cells of the same compositions. The
latter model random alloys wherein the Mg and Ti are
distributed over the lattice sites of a TiH$_2$-like structure.
Details of these models are presented in
Ref.~\onlinecite{er07}.

The dielectric functions consist of intra and interband contributions:
\begin{equation}
\varepsilon(\omega)=\varepsilon_{\rm inter}(\omega)+\varepsilon_{\rm intra}(\omega)~~.
\label{dieladdi}
\end{equation}
These are calculated separately. For the interband contribution
$\varepsilon_{\rm inter}$ we use first-principles DFT in the
independent particle random phase approximation. Since the
materials in question are metals, the intraband dielectric part
$\varepsilon_{\rm intra}$ does not vanish. It is modeled from
the plasma frequencies $\omega_{\textrm{p}}$, which are
calculated from first-principles as well.

In Sec.~\ref{compu} the technical details of the calculations
are summarized. Sec.~\ref{struclab} contains a concise
description of the structural models used. Results on the
dielectric functions are presented in Sec.~\ref{results}.
Finally, in the discussion section (Sec.~\ref{discuss}) results
are compared with experiment.\cite{borsa07}

\section{Computational methods}
\label{compu}

First-principles DFT calculations were carried out with the
Vienna \em Ab initio \em Simulation Package
(VASP),\cite{vasp1,vasp2,vasp3} using the projector augmented
wave (PAW) method.~\cite{paw,blo} A plane wave basis set was
used and periodic boundary conditions applied. The kinetic
energy cutoff on the Kohn-Sham states was 312.5~eV. For the
exchange-correlation functional we used the generalized
gradient approximation (GGA).\cite{gga} Non-linear core
corrections were applied.\cite{core}

The Brillouin zone integrations were performed using a Gaussian
smearing method with a width of 0.1~eV.\cite{blochl94}. The
$\bf k \rm$-point meshes were even grids containing $\Gamma$ so
that the band extrema are typically included in the calculation
of the dielectric functions. The convergence of the dielectric
functions and intraband plasma frequencies with respect to the
$\bf k \rm$-point meshes was tested by increasing the number of
$\bf k \rm$-points for each system separately. A typical mesh
spacing of about 0.01~\AA$^{-1}$\ was needed to obtain
converged results.

The calculations of the complex interband dielectric functions,
$\varepsilon_{\rm inter}(\omega) = \varepsilon_{\rm
inter}^{(1)}(\omega) + i \varepsilon_{\rm
inter}^{(2)}(\omega)$, were performed in the random phase
independent particle approximation, i.e.\ taking into account
only direct transitions from occupied to unoccupied Kohn-Sham
orbitals. Local field effects were neglected. The imaginary
part of the macroscopic dielectric function, $\varepsilon_{\rm
intra}^{(2)}(\omega)$, than has the form:
\begin{eqnarray}\label{epsinter}
\varepsilon_{\rm inter}^{(2)}(\mathbf{\hat{q}},\omega)&=&\frac{8\pi^2 e^2}{V}
\limq \frac{1}{\abs{\bq}^2}\times\\\nonumber
&& \sum_{\bk,v,c} \abs{\bracet{u_{c,\bk + \bq}}{u_{v,\bk}}}^2
\delta(\epsilon_{c,\bk + \bq}-\epsilon_{v,\bk}-\hbar\omega)
\end{eqnarray}
where $\mathbf{\hat{q}}$ denotes the direction of $\bq$ and
$v,\mathbf{k}+\mathbf{q}$ and $c,\mathbf{k}$ label single
particle states that are occupied and unoccupied in the ground
state, respectively. $\epsilon$, $u$ are the single particle
energies and the translationally invariant parts of the wave
functions. $V$ is the volume of the unit cell. The real part,
$\varepsilon_{\rm inter}^{(1)}(\omega)$, is obtained via a
Kramers-Kronig transform. Further details can be found in
Ref.~\onlinecite{kresseps}.

Most optical data on hydrides are obtained from micro- and
nano-crystalline samples whose crystallites have a significant
spread in orientation. Therefore the most relevant quantity is
the directionally averaged dielectric function, i.e.\
$\varepsilon_{\rm inter}^{(2)}(\omega)$ averaged over
$\mathbf{\hat{q}}$.

\begin{figure}[!tbp]
\centering
\includegraphics[angle=270,width=8.5cm]{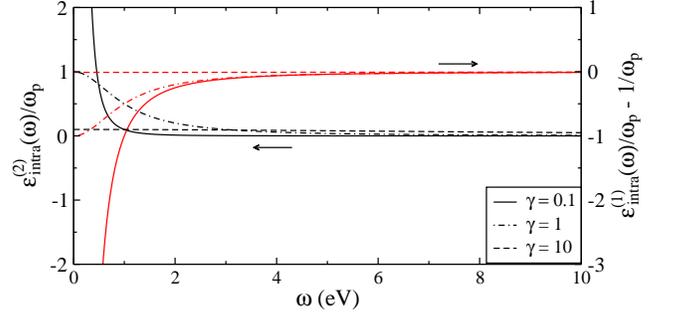}
\caption{\label{epsintraplot}(Color online) Dielectric functions of a free electron gas according to equation~\ref{epsintra}.}
\end{figure}

The intraband dielectric function, $\varepsilon_{\rm
intra}(\omega) = \varepsilon_{\rm intra}^{(1)}(\omega) + i
\varepsilon_{\rm intra}^{(2)}(\omega)$, is calculated from the
free electron plasma frequency $\omega_{\textrm{p}}$:
\begin{eqnarray}\label{epsintra}
\varepsilon_{\rm intra}^{(1)}(\omega) &=&  1 - \frac{\omega^2_{\textrm{p}}}{\omega^2 + \gamma^2}\\
\varepsilon_{\rm intra}^{(2)}(\omega) &=&  \frac{\gamma \omega^2_{\textrm{p}}}{\omega^3 + \omega  \gamma^2}
\end{eqnarray}
where an inverse lifetime, $\gamma = 0.01$~eV, is used. For
$\gamma=0$ the reflection would be perfect up to the plasma
frequency and zero beyond. Finite values of $\gamma$ decrease
the reflection below $\omega_{\textrm{p}}$ and smoothen the
reflection edge at $\omega_{\textrm{p}}$. For metals $\gamma$
values are in the order of 0.1~eV. By using a small value the
influence of the interband part is emphasized. Calculating
$\gamma$ from first principles obviously goes beyond DFT. For
three values of $\gamma$ the free electron $\varepsilon_{\rm
intra}$ is plotted in Fig.~\ref{epsintraplot}.

The plasma frequency $\omega_{\textrm{p}}$ is calculated as an
integral over the Fermi surface according to:
\begin{equation}\label{plasmafreq}
\omega^2_{\textrm{p} (\alpha \beta)} = \frac{4\pi e^2}{V \hbar^2} \sum_{n,\mathbf{k}} 2 g_{\mathbf{k}}\frac{\partial f(\epsilon_{n\mathbf{k}})}{\partial \epsilon}\left(\mathbf{e}_\alpha \frac{\partial \epsilon_{n\mathbf{k}}}{\partial \mathbf{k}}\right)\left(\mathbf{e}_\beta \frac{\partial \epsilon_{n\mathbf{k}}}{\partial \mathbf{k}}\right)
\end{equation}
with $g_{\mathbf{k}}$ the weight factors, and
$f(\epsilon_{n\mathbf{k}})$ the occupation function. Again we
use directionally averaged values. Further details on the
calculation of the plasma frequency can be found in
Ref.~\onlinecite{kressplasma}.

Finally the optical constants, the refraction index, $n$, and
the extinction coefficient, $\kappa$, and the absorption, $A$,
reflection, $R$, and transmission, $T$, are calculated using
the standard expressions:
\begin{eqnarray}\label{nkart}
n&=&\frac{1}{\sqrt{2}}\sqrt{\varepsilon^{(1)}+\sqrt{{\left( \varepsilon^{(1)} \right)}^2+ \left( \varepsilon^{(2)} \right)^2}}\\
\kappa&=&\frac{1}{\sqrt{2}}\sqrt{-\varepsilon^{(1)}+\sqrt{ \left( \varepsilon^{(1)} \right)^2 + \left( \varepsilon^{(2)} \right)^2}}\\
A&=&1 - \exp(-\kappa\omega d / c)\\
R&=&\frac{(n-1)^2+\kappa^2}{(n+1)^2+\kappa^2}\\
T&=&(1-R)(1-A),
\end{eqnarray}
with $\varepsilon^{(1)}$ and $\varepsilon^{(2)}$ the real and
imaginary part of $\varepsilon$, $d$ the slab thickness and $c$
the speed of light in vacuum. The reflection and transmission
spectra are constructed to simulate the
substrate/Mg$_x$Ti$_{(1-x)}$H$_2$/palladium setup as was used
in the experiments by Borsa {\it et~al.}\cite{borsa06,borsa07}
All internal reflections and absorptions in the three layer
system are taken into account.

\section{Structures}
\label{struclab}

The calculation of the dielectric functions is performed using
the crystal structures developed in Ref.~\onlinecite{er07}. A
brief summary of compositions and cell parameters is given in
Table~\ref{structable}. In short they were constructed in the
following way.

In the case of $x=0.875$ the simple cell is just the optimized
experimental cell with composition Mg$_{28}$Ti$_4$H$_{64}$. For
$x=0.5$ and 0.75, two and three atoms, respectively, out of the
four titanium atoms in the conventional fcc TiH$_2$ cell were
replaced by magnesium. Thus the unit cells have compositions
Mg$_2$Ti$_2$H$_8$ and Mg$_3$TiH$_8$ respectively.

To simulate the random alloys super cells are used. These are
also based on the fluorite Ti$_4$H$_8$ ($Z=1$) building block.
For $x=0.5$ and 0.75 $2\times2\times2$ super cells were
constructed and for $x=0.875$ a $2\times1\times1$ super cell.
Again Ti were replaced by Mg, but now such as to approximate
random alloys most efficiently (see Ref.~\onlinecite{er07} for
details).

For all the models constructed the positional and cell
parameters were relaxed.
The cells remain close to cubic, see
Table~\ref{structable}. The angles between the crystal axes are
close to 90$^\circ$, except for the $x=0.75$ and 0.875 super
cells where there is a small deviation.

\begin{table}[!tbp]
\caption{Number of formula units (Z), lattice parameters (\AA) and shortest Ti--Ti interatomic distances (\AA). The first row for each composition contains the data of the simple cell and the second of the super cell.\label{structable}}
\begin{ruledtabular}
\begin{tabular}{lcrrrr}
x     & Z  &    a &    b &    c & Ti--Ti \\
\hline
0.5   & 4  & 4.72 & 4.72 & 4.65 & 3.16   \\
      & 32 & 9.08 & 9.03 & 9.09 & 3.15   \\
0.75  & 4  & 4.62 & 4.62 & 4.62 & 4.62   \\
      & 32 & 9.29 & 9.30 & 9.27 & 3.13   \\
0.875 & 32 & 9.36 & 9.36 & 9.36 & 6.62   \\
      & 64 &18.75 & 9.42 & 9.73 & 3.08   \\
\end{tabular}
\end{ruledtabular}
\end{table}


\section{Dielectric functions}
\label{results}

\subsection{Interband dielectric functions}
\label{intersub}

\begin{figure}[!tbp]
\centering
\includegraphics[angle=270,width=8cm]{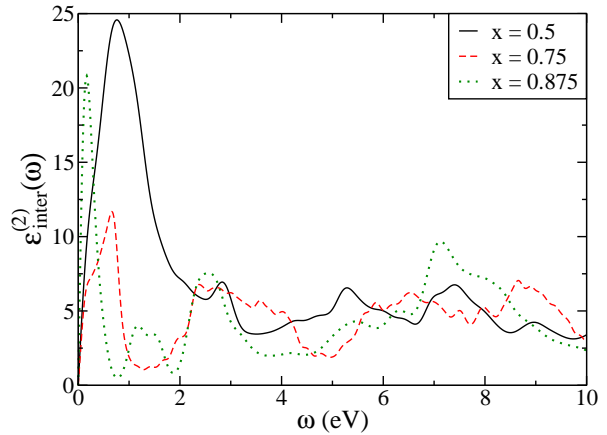}
\caption{\label{epssimple}(Color online) Imaginary parts of the interband dielectric functions of Mg$_x$Ti$_{(1-x)}$H$_2$ in the primitive cells.}
\end{figure}

\begin{figure}[!tbp]
\centering
\includegraphics[angle=270,width=8cm]{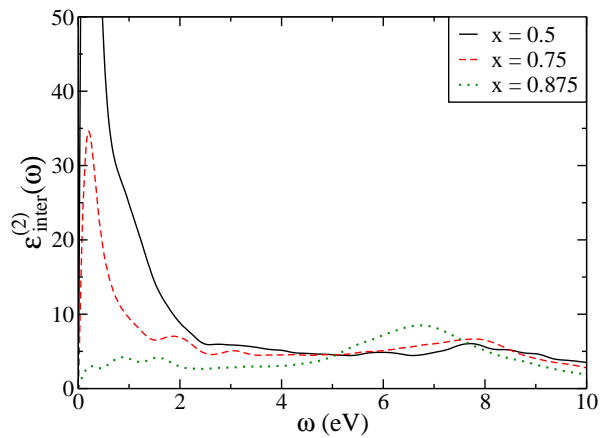}
\caption{\label{epssuper}(Color online) Imaginary parts of the interband dielectric functions of Mg$_x$Ti$_{(1-x)}$H$_2$ in the super cells. For $x = 0.5$ the dielectric function peaks to a value of 130.}
\end{figure}

Figures \ref{epssimple} and \ref{epssuper} show the calculated
imaginary parts of the interband dielectric functions in the
simple and super cells respectively. In general the dielectric
functions exhibit a peak at low energy, below 2~eV, followed by
a relatively flat tail. In the super cells the dielectric
functions have higher peaks and flatter tails. An exception is
the dielectric function of the $x=0.875$ super cell. It does
not have a peak at low energy.

\begin{figure}[!tbp]
\centering
\includegraphics[width=8cm]{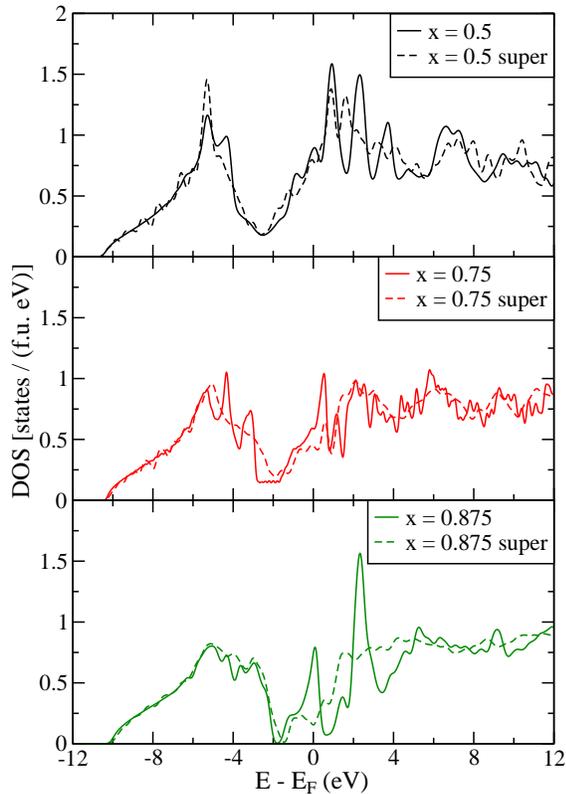}
\caption{\label{dossen}(Color online) Electronic densities of state of the simple and super cells for the three compositions of Mg$_x$Ti$_{(1-x)}$H$_2$. The zero of energy is taken at the Fermi level.}
\end{figure}

\begin{figure}[!tbp]
\centering
\includegraphics[width=8cm]{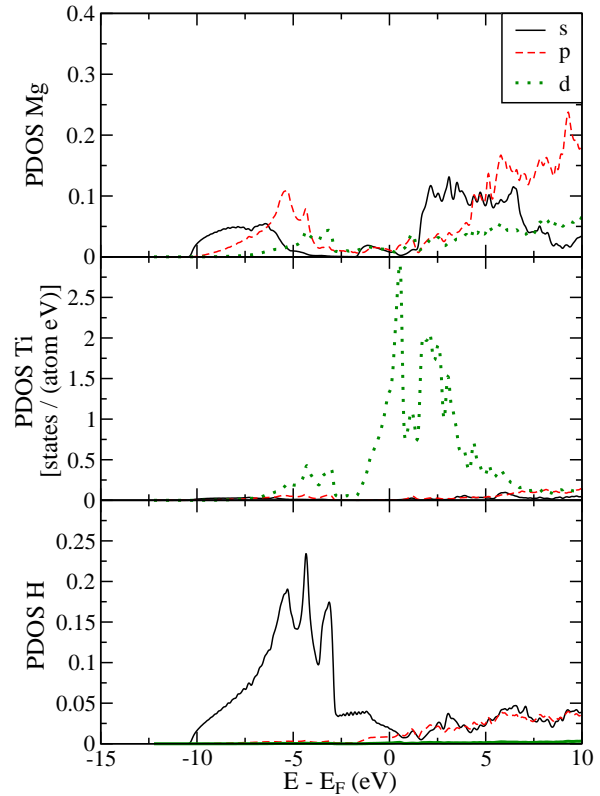}
\caption{\label{pdos}(Color online) Angular momentum projected partial densities of states (PDOS) [states/(atom~eV)] of Mg$_{0.75}$Ti$_{0.25}$H$_2$ in the simple unit cell. The zero of energy is
at the Fermi level. The PDOS' are calculated in spheres with radii of 1.3, 1.3 and 0.8~\AA\ for Mg, Ti an H respectively.}
\end{figure}

To understand the main features of the dielectric functions we
study the densities of states (DOS', Fig.~\ref{dossen}). Below
the Fermi level the DOS' have an approximately triangular shape
with a predominant hydrogen-$s$ and magnesium-$s$ and
$p$~character. At the Fermi level the DOS' are dominated by
titanium-$d$ states. Above the Fermi level the DOS' have a
mixed character of hydrogen-$s$ and $p$, magnesium-$s$ and $p$
and titanium-$d$. As an illustration the angular momentum
projected partial DOS' of the simple cell with $x = 0.75$ is
shown in Fig.~\ref{pdos}.

\begin{figure}[!tbp]
\centering
\includegraphics[width=8cm]{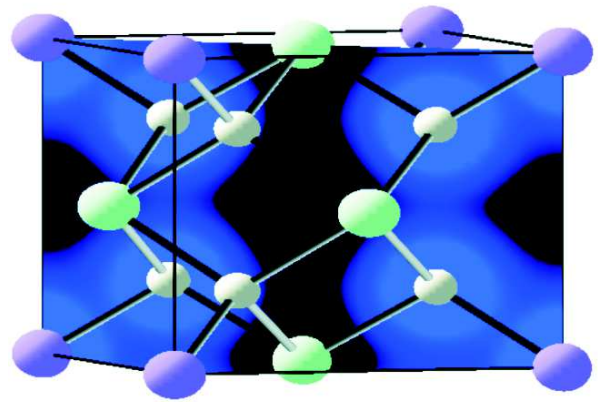}
\caption{\label{struc}(Color online) Electron localization function of Mg$_{0.75}$Ti$_{0.25}$H$_2$ in the simple unit cell. The Ti atoms are at the corners, the Mg atoms at the face centers and the H atom (small) are inside the cell.
The electron localization function is plotted on a (110) cut-plane.
Localization at the H and Ti atoms is evident.}
\end{figure}

The electron density is localized at both the hydrogen and
titanium atoms. This is illustrated in Figure~\ref{struc},
where the electron localization function
(Ref.~\onlinecite{silvi94}) is plotted, also for the simple
cell Mg$_{0.75}$Ti$_{0.25}$H$_2$. At the titanium atoms the
only available states are of $d$-character. Since $d-d$
transitions are very weak the main contribution to the
dielectric function is from $s-p$ transitions on the hydrogen
atoms.

In first order the imaginary part of the dielectric function
can be described as the joint density of states (JDOS) divided
by $\omega^2$. The hydrogen DOS has a dip near the Fermi
energy. It increases when moving away from the Fermi energy,
both to lower and higher energies. This causes the JDOS at the
hydrogen atoms to increase more than linearly (in the region
from 0 to about 8~eV). When divided by $\omega^2$ this increase
is rather effectively compensated. Hence the dielectric
function does not vary much in the interval from 2 to 8~eV.
This reasoning applies for all titanium concentrations that we
considered. In the super cells the averaging over the various
hydrogen DOS' gives rise to a further leveling out of the
dielectric functions. Of course vibrational effects, that are
lacking in our 0~K calculation, will even further smoothen the
dielectric functions.

The peaks at the lower end of the energy range do arise from
d-d transitions on the Ti atoms. Although their oscillator
strength is small, division by $\omega^2$ causes them to stand
out nevertheless. For $x=0.875$ the Ti DOS at $E_{\rm Fermi}$
is strongly suppressed in the super cell (see
Fig.~\ref{dossen}). This correlates with the absence of the
peak in $\epsilon_{\rm inter}^{(2)}$ for the super cell at this
composition (Figs.~\ref{epssimple},~\ref{epssuper}). At the
other two compositions, we see no clear correlation between
difference in DOS and peak shape (comparing simple and super
cells). The higher peaks in the DOS in the super cells can be
understood when we realize that the effective ``back folding''
of the d bands {\em and} their mutual interactions (caused by
the randomization) results in a flattening of the bands. Some
of these bands will be very close to $E_{\rm Fermi}$ and their
transitions will thus be ``boosted'', both by the flatness of
the bands and the small transition energy. This discussion
anticipates the discussion in the next section.

\subsection{Intraband plasma frequencies}

The intraband plasma frequencies, which have been calculated
according to Equation~\ref{plasmafreq}, are listed in
Table~\ref{plasma}. The squared plasma frequencies from the
super cell calculations are between one and two orders of
magnitude smaller than those of the simple cells. Therefore the
edge on the free carrier reflectively occurs at considerably
lower energies in the random alloys. In our models the maximum
$\omega_{\rm p}$ is 1.1~eV. This is, however, for only one
realization of a random model at $x=0.5$ and it is well
conceivable that calculations on large models could yield even
lower $\omega_{\rm p}$. For the simple cells, the plasma
frequencies are approximately 3~eV, hence these systems are
highly reflecting for $\omega < 3$~eV.

\begin{table}[!tbp]
\caption{Squared plasma frequencies, $\omega_{\textrm{p}}^2$
(eV$^2$), from intraband transitions.\label{plasma}}
\begin{ruledtabular}
\begin{tabular}{lcc}
               & simple cell & super cell \\
\hline
$x = 0.5$      &        10.7 &        1.3 \\
$x = 0.75$     &        12.8 &        0.4 \\
$x = 0.875$    &         8.6 &        0.1 \\
\end{tabular}
\end{ruledtabular}
\end{table}

Eq.~\ref{plasmafreq} is a good starting point for a discussion
of the trends across Table~\ref{plasma}. It basically states
that the squared plasma frequency is proportional to the
product of the electron density and the square of the slope of
the energy bands, both calculated at the Fermi level. In the
super cells $\omega_{\textrm{p}}^2$ clearly follows the density
of states (see, Fig.~\ref{dossen}) as a lower concentration of
titanium means a lower amount of free electrons. In the simple
cells this trend is not obvious. Subtle changes in the shape of
the DOS, i.e.\ bandstructure effects, make for less changes in
the electron density at the Fermi level. Consistently the
plasma frequency shows little variation. This, however, does
not imply that small changes in the slope of the energy bands
do not play an equally important role here.

\begin{figure}[!tbp]
\centering
\includegraphics[width=8cm]{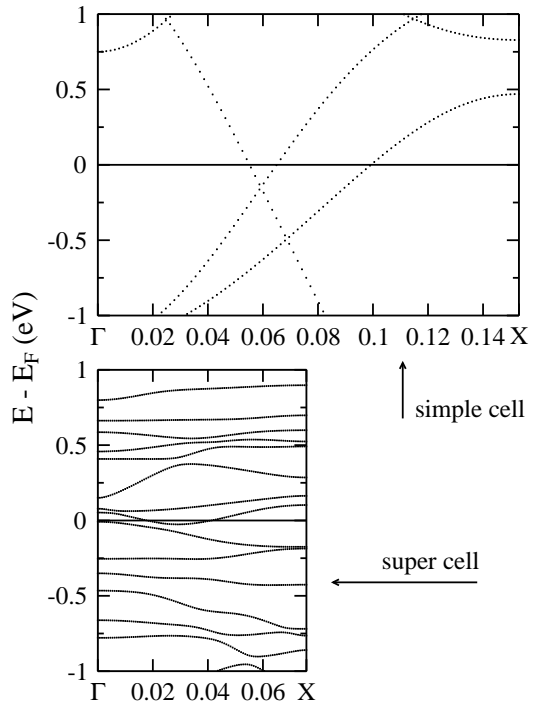}
\caption{\label{bands}Electronic energy band structure of Mg$_{0.75}$Ti$_{0.25}$H$_2$ in the simple and super cell along the $\Gamma$--X direction.}
\end{figure}

The difference between the simple and super cells of the same
composition is explained by a decrease in energy band slope.
The DOS' at the Fermi level are of comparable size, at least
for compositions with $x=0.5$ and $x=0.75$. This clearly points
to the change in $\partial \epsilon_{n \bf k}/\partial
\epsilon_{\bf k}$ as the cause of the significant decrease of
$\omega_{\rm p}$. This is corroborated by inspection of the
band structure, that is plotted in Fig.~\ref{bands} for
Mg$_{0.75}$Ti$_{0.25}$H$_2$ for both the simple and super cell.
We see that the bandstructure of the supercell cannot be
understood as a simple back-folding of the bands. The
randomness in the structure has induced many interactions
between the $d$-bands, leading to a dramatic reduction of their
slopes. Going to even larger super cells, the effect may become
even stronger, and the free carrier reflection concomitantly
reduced. Such a calculation goes beyond the present study. It
may require a different formalism as the reciprocal space
description of Eq.~\ref{plasmafreq} is bound to break down for
truly random systems.

Interestingly, the flatness of the energy bands also makes that
the effective mass of the ``free'' electrons is rather high.
This goes some way to explain the rather high resistivity
measured for these systems. The ``flat bands'' also point to a
possible localization of carriers.

For $x=0.875$ the reduction of $\omega_{\textrm{p}}^2$ cannot
be understood as a reduction of only $\partial \epsilon_{n \bf
k}/\partial {\bf k}$. From Fig.~\ref{dossen} a substantial
reduction of the DOS at $E_{\rm Fermi}$ is evident. Hence
reduction of $\omega_{\textrm{p}}^2$ is much stronger than for
$x=0.5$ and $x=0.75$.

To obtain the dielectric functions of the materials we just add
the interband part of Sect.~\ref{intersub} and intraband parts
obtained from $\omega_{\textrm{p}}$ according to
Eq.~\ref{dieladdi}. It was already noted above that the impact
on the reflection in the visible range will be substantial for
the simple cells. The values of $\omega_{\textrm{p}}^2$ in the
super cells are low enough to only induce minor corrections to
the interband dielectric function. The implications of the
corrections will be discussed in the next Section.


\section{Discussion}
\label{discuss}

The calculations on the random alloy super cells clearly
demonstrate that breaking of the short range order results in
two important effects: (a) The interband dielectric function is
smoothened and (b) the plasma frequency is lowered. The first
translates in smoother reflection and transmission spectra for
the super cells. The second results in a lower reflection edge.
The simple cells have almost full reflection up to 1--2~eV
whereas in the super cells full reflection only occurs below
$\sim$0.3~eV.

\begin{figure}[!tbp]
\centering
\includegraphics[width=8cm]{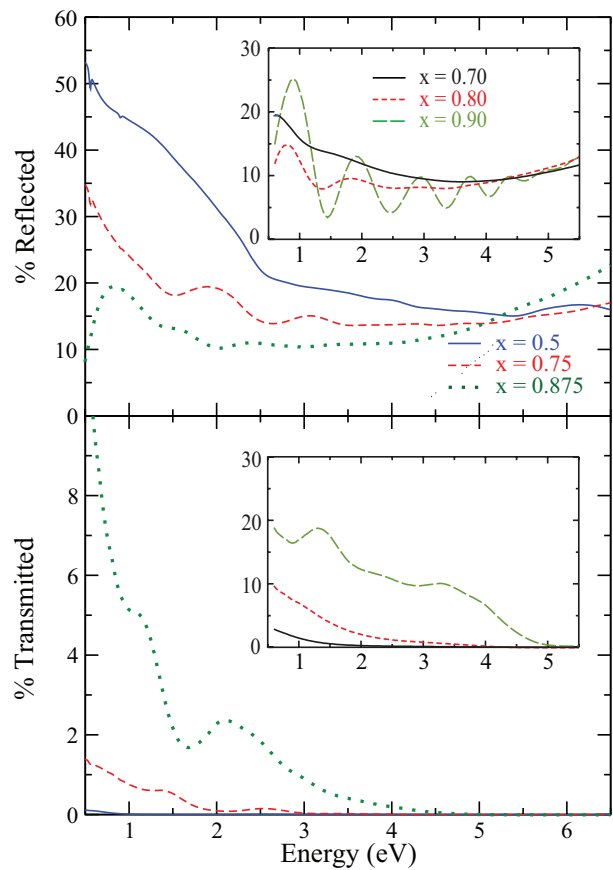}
\caption{\label{rt}(Color online) Reflection and transmission of a
10~nm Paladium / 200~nm Mg$_x$Ti$_{(1-x)}$H$_2$ / $(k = 0)$ quartz substrate film.
For the dielectric function of Mg$_x$Ti$_{(1-x)}$H$_2$ the results for the supercell,
including inter and intraband contributions, are used. The inset shows experimental
results from Ref.~\onlinecite{borsa07}, reproduced with permission, copyright American
Physical Society (2007).}
\end{figure}

Figure~\ref{rt} shows the calculated reflection and
transmission of a film simulating closely the experimental
setup as used by Borsa {\it et al.}\cite{borsa07} The simulated
film consists of 10~nm Paladium / 200~nm
Mg$_x$Ti$_{(1-x)}$H$_2$ / $(k = 0)$ quartz substrate layers,
were we use the $\varepsilon(\omega)$ calculated in the super
cells for the Mg$_x$Ti$_{(1-x)}$H$_2$ layer. The plotted range
is that of a Perkin Elmer Lambda 900 diffraction spectrometer.
For comparison; visible light lies in the range of
1.65--3.26~eV. The general features and trends as a function of
Ti content in both reflection and transmission are in agreement
with the trends observed in the  experiments.\cite{borsa07}
Indeed the reflection is low and our calculations describe a
``black state''. With decreasing Ti content the reflection
decreases and the transmission increases.

Since experiments were performed on magnesium/titanium ratios
$x=0.7$, 0.8, and 0.9, that would have required very large
super cells, a detailed comparison can only be made partially.
Interpolating between the experimental values for 0.7 and 0.8
we can make a reasonable comparison with the calculated results
for $x=0.75$. The experimental reflection lies about 5 \%Pt.\
lower than the calculated one but the shape of the curve is
very similar. The experimental transmission drops from
6.5\%--0\% in the range from 0.5--3~eV, it hence lies about a
factor 3 higher than the calculated one. The shapes of the
transmission curves, however shows good agreement.

Comparing the calculated values for $x=0.875$ to the
experimental results for $x=0.9$ we again see somewhat lower
values for the experimental reflection. Furthermore, the slight
oscillation seen in the calculated spectrum, caused by
interference, is much stronger in the experimental spectrum.
The main difference in the transmission lies in the energy
range below 1~eV where the calculated transmission is much
larger than the experimental values. This difference could
point to an underestimation of the $\omega_{\textrm{p}}^2$ in
the super cell $x=0.875$ calculation.

From the correspondence between the experimental and calculated
optical spectra we conclude that the ``black state'' can
already be explained from moderate-size randomized super cells.
Put into other words: randomized models containing only 32 to
64 formula units allow for a breaking of the order on a
length-scale such as to lower the reflection edge and smoothen
the spectra. This does not necessarily imply that a
randomization at larger length scales, with a concomitant
increase of the short range order, e.g.\ in the coherent
crystal picture proposed by Borsa {\it et al.}\cite{borsa07},
would be inconsistent with experiment. Indeed, the coherent
crystal model is supported by various observations, e.g.\ the
large positive enthalpy of mixing of Ti and Mg. Moreover, for
$x=0.875$ our modeling seems to underestimate $\omega_{\rm p}$,
suggesting that the short range disorder may have disrupted the
band dispersion to much. A full first-principles study of
necessarily larger models with more short range order is beyond
present computational capabilities. Such a study is desirable.
However, with the present randomized models we can capture most
of the essential physics of Mg$_x$Ti$_{1-x}$H$_2$.

\begin{acknowledgments}
The authors thank Prof.\ G.\ Kresse and J.\ Harl
(Universit{\"a}t Wien) for the use of the optical packages and
R. Gremaud (Vrije Universiteit) for running the
reflection/transmission simulations. This work is part of the
Sustainable Hydrogen Programme of the Advanced Catalytic
Technologies for Sustainability (ACTS) and the Stichting voor
Fundamenteel Onderzoek der Materie (FOM), both financially
supported by the Nederlandse Organisatie voor Wetenschappelijk
Onderzoek (NWO).
\end{acknowledgments}

\bibliography{proefschrift}

\end{document}